\documentclass[%
reprint, twocolumn,
superscriptaddress,
showpacs,preprintnumbers,
nofootinbib,
amsmath,amssymb, aps, pra,
]{revtex4}

\usepackage{graphicx}
\usepackage{dcolumn}
\usepackage{bm}
\usepackage{color} 
\usepackage{CJK}
\usepackage{dsfont}
\usepackage[extension=xxx]{hyperref}

\newcommand{\beq}{\begin{equation}}
\newcommand{\eeq}{\end{equation}}
\newcommand{\beqa}{\begin{eqnarray}}
\newcommand{\eeqa}{\end{eqnarray}}

\def\beq{\begin{equation}}

\begin{document}

\title{Fast driving  between arbitrary states of a quantum particle by trap deformation}
\author{S. Mart\'\i nez-Garaot}
\email{sofia.martinez@ehu.eus}
\affiliation{Departamento de Qu\'{\i}mica F\'{\i}sica, UPV/EHU, Apdo
644, 48080 Bilbao, Spain}
\author{M. Palmero}
\affiliation{Departamento de Qu\'{\i}mica F\'{\i}sica, UPV/EHU, Apdo
644, 48080 Bilbao, Spain}
\author{J. G. Muga}
\affiliation{Departamento de Qu\'{\i}mica F\'{\i}sica, UPV/EHU, Apdo
644, 48080 Bilbao, Spain}
\author{D. Gu\'ery-Odelin}
\affiliation{Laboratoire de Collisions Agr\'egats R\'eactivit\'e, CNRS UMR 5589, IRSAMC, Universit\'e de Toulouse (UPS), 118 Route de Narbonne, 31062 Toulouse CEDEX 4, France}

\date{\today}
\begin{abstract}
By performing a slow adiabatic change between two traps of a quantum particle, 
it is possible to transform an eigenstate of the original trap into the 
corresponding eigenstate of the final trap.
If no level crossings are involved, the process can be  made faster than adiabatic by setting first the interpolated 
evolution of the wave function from its initial to its final form and inferring from this evolution the trap deformation. 
We find a simple and compact formula which gives the trap shape at any time for any interpolation scheme. It is applicable 
even in complicated scenarios where  there is no adiabatic process 
for  the desired state-transformation, e.g., if the state changes its topological properties. 
We illustrate its use for the expansion   
of a harmonic trap, for the transformation of a harmonic trap into a linear trap and into an arbitrary number of traps of a
periodic structure. Finally, we study the creation of a node
exemplified by the passage from the ground state to the first excited state of a harmonic oscillator.  
\end{abstract}
\maketitle
\section{Introduction}
There is a growing interest in accelerating transformations among different quantum states to limit the detrimental effect of decoherence or noise, or to increase the repetition rate, or the number of quantum operations that can be carried out in a given time interval \cite{reviewSTA2013}. In this context, a few methods that bypass adiabatic transformations by designing 
appropriate time-dependent Hamiltonians have been set up. This includes methods based on exact solutions with time-dependent scaling parameters \cite{DGO2009,PRL10,boltz14,PRA2014}, methods based on dynamical Lewis-Riesenfeld invariants \cite{PRL10,NJP2012,PRL13}, the transitionless tracking algorythm that adds counteradiabatic terms to the Hamiltonian \cite{Rice,Berry}, the fast-forward approach \cite{ffw1,ffw2,torron12,torron13}, Lie-algebraic methods \cite{Lie1,Lie2}, or the fast quasi-adiabatic approach \cite{FAQUAD}.

Recent experimental progress enables one to shape atomic traps dynamically using for instance lasers diffracted by Spatial Light Modulators (see e.g. \cite{slm}), time-dependent microwave dressing \cite{rfdressing}, or, for ions in multisegmented Paul traps,  time-dependent voltages applied to the control electrodes \cite{Niza}. Such trap shaping on a short timescale has been proven useful, for example, 
to implement quantum thermodynamical cycles \cite{quantumthermo1,quantumthermo2}, or to implement a scalable architecture for quantum information processing
\cite{Kielpinski}. It is also important for some quantum information processing schemes such as multiplexing and demultiplexing when information is encoded in external degrees of freedom \cite{PRL13}, for Fock state creation \cite{Fock}, or velocity control \cite{DMR,GM},   
and it is expected to become more and more relevant given the current interest to develop
quantum technologies. 

In this article, we design the fast driving of a wave packet for a particle in a time and position-dependent trap potential with the aim of reaching some target state. We provide an explicit formal solution for the time-dependent potential that connects eigenstates of different traps, or different eigenstates 
of the same trap. The theoretical framework developed here builds on, and improves, the one presented in \cite{torron12,torron13}, where a streamlined 
version of the fast-forward method of Masuda and Nakamura \cite{ffw1,ffw2} was derived and exemplified to drive a matter wave from a single well to a symmetric double-well. The solution presented here reduces considerably the numerical time necessary to generate the appropriate time-dependent potential, and offers therefore the possibility to explore more complex state transformations. It also enables us to drive transitions that could not be handled with the techniques in  \cite{torron12,torron13}, specifically transformations that change the topology of the state with the creation of a node. 

In  Sec.~\ref{sec2} we introduce the theoretical framework and explain the inverse protocol procedure. Section ~\ref{ex}  presents several examples including ground state to ground state and first-excited to first-excited transformations, matter wave splitting in an arbitrary number of traps, and the transformation between states with different topology. Finally, in Sec.~\ref{sec4} we discuss the results and open questions.
\section{From wave function to potential}
\label{sec2}
To drive a wave function from an eigenstate of an initial potential to an eigenstate of a final potential (which might be identical to the 
original one),  
we use a similar approach to the one 
in \cite{torron12,torron13}.  
In this approach, the initial wave function, $\psi_{\rm i}(x)$, and final (target) wave function, $\psi_{\rm f}(x)$,  are given. 
The wave function should evolve between these two states in a predetermined time $t_{\rm f}$, satisfying the boundary conditions $\psi(x,0)=\psi_{\rm i}(x)$ and $\psi(x,t_{\rm f})=\psi_{\rm f}(x)$. The corresponding time-dependent potential can in principle be deduced from the Schr\"odinger equation, 
\begin{equation} 
V(x,t)= \frac{1}{\psi(x,t)} \left(\displaystyle i\hbar\frac{\partial \psi(x,t)}{\partial t} + \frac{\hbar^2}{2m}\frac{\partial ^2 \psi(x,t)}{\partial x^2} \right).
\label{eq1}
\end{equation}
Using the modulus-phase representation for the time-dependent wave function, 
\beq
\label{m-p}
\psi(x,t)=\rho(x,t)e^{i\phi(x,t)},
\eeq
the expression for the potential can be worked out, which in general becomes a complex function with real and imaginary parts. 
By imposing that the potential takes real values, ${\rm Im}(V(x,t))=0$, we get a first relation that links the phase $\phi(x,t)$ and the modulus $\rho(x,t)$,
\begin{equation} 
\frac{1}{\rho}\frac{\partial \rho}{\partial t} + \frac{\hbar}{2m}\left( \frac{2}{\rho}\frac{\partial \phi}{\partial x}\frac{\partial \rho}{\partial x}  + \frac{\partial^2 \phi}{\partial x^2}  \right)=0.
\label{eq2}
\end{equation}
This is a continuity equation. The expression for the potential then reads
\begin{equation} 
V(x,t) = -\hbar \frac{\partial \phi}{\partial t} + \frac{\hbar^2}{2m}\left[  \frac{1}{\rho}\frac{\partial^2 \rho}{\partial x^2} - \left( \frac{\partial \phi}{\partial x}  \right)^2  \right].  
\label{eq3}
\end{equation}
Equation~(\ref{eq2}) can be integrated formally,
\begin{equation} 
\frac{\partial \phi}{\partial x} = - \frac{mu(x,t)}{\hbar},
\label{eq4}
\end{equation}
where $u$ plays the role of a  ``hydrodynamic velocity'',
\begin{equation} 
u(x,t) = \frac{1}{\rho^2(x,t)} \frac{\partial}{\partial t} \left( \int_0^x \rho^2(x',t)dx' \right).
\label{eq5}
\end{equation}
The potential $V(x,t)$ can therefore be inferred  from $\rho(x,t)$,
\begin{eqnarray} 
V(x,t)  & = &    m\frac{\partial }{\partial t} \int_0^x u(x',t)dx' +
\frac{\hbar^2}{2m}  \frac{1}{\rho(x,t)}\frac{\partial^2 \rho (x,t)}{\partial x^2} 
\nonumber \\  & - &  \frac{1}{2} mu^2(x,t) -\hbar \dot \phi_0(t),
\label{eqpot}
\end{eqnarray}
where $\phi_0\equiv\phi(x=0,t)$ and the dot means time derivative.
This is our central result. 

In a previous streamlined version of the fast-forward approach \cite{torron12,torron13}, $\rho(x,t)$ was designed first and then the equation for the phase had to be solved numerically in order to get the potential. However, in the current improved formulation
the potential only depends on $\rho(x,t)$, so we find it directly from Eq. (\ref{eqpot}). This is a great advantage because the computation time necessary to generate $V(x,t)$ is reduced considerably. 
Furthermore, as we will see in the examples, by reinterpreting the modulus and phase decomposition 
we can also solve more complicated problems. 

Consider a transformation  from the ground state, $\psi_{\rm i}(x)$, of a potential $V(x,0)$, to the ground state, $\psi_{\rm f}(x)$, of another potential, $V(x,t_{\rm f})$
in a time $t_{\rm f}$. 
We may use the interpolation formula
\begin{equation} 
\rho(x,t) = {\cal N}(t) \left\{  \left[1-\eta(t)\right]\rho_{\rm i}(x) + \eta(t)\rho_{\rm f}(x) \right\}, 
\label{rho_p}
\end{equation}
where $\rho_{\rm i}(x)=|\psi_{\rm i}(x)|$ and $\rho_{\rm f}(x)=|\psi_{\rm f}(x)|$, 
$\eta(t)$ is a monotonous and smooth function that varies from $\eta(0)=0$ to $\eta(t_{\rm f})=1$, and ${\cal N}(t)$ is a normalization factor. 
As the wave functions are ground states they have no nodes and the quantity $\rho(x,t)$ never vanishes. This ensures the absence of any divergent behavior in the potential (\ref{eqpot}).  
Furthermore, $\rho(x,t)$ in Eq. (\ref{rho_p}) is positive during the whole process by construction. 

Assuming  the boundary conditions 
\beqa
\label{b_c}
\dot \eta(0)&=&\dot \eta(t_{\rm f})=0,
\nonumber\\ 
\ddot \eta(0)&=&\ddot \eta(t_{\rm f})=0,
\eeqa
we find  from Eqs. (\ref{eq5}) and (\ref{rho_p}),
\beqa
\label{b_c_ru}
&&\dot\rho(x,0)=\dot\rho(x,t_f)=0,
\nonumber \\
&&u(x,0)=u(x,t_f)=0,
\eeqa
and consequently,
\begin{eqnarray} 
V(x,0) & = & -\hbar \dot \phi_0(0) + \frac{\hbar^2}{2m}  \frac{1}{\psi_{\rm i}(x)}\frac{\partial^2 \psi_{\rm i}}{\partial  x^2},
\nonumber \\
V(x,t_{\rm f}) & = & -\hbar \dot \phi_0(t_{\rm f}) + \frac{\hbar^2}{2m}  \frac{1}{\psi_{\rm f}(x)}\frac{\partial^2 \psi_{\rm f}}{\partial x^2}. 
\label{eq8}
\end{eqnarray}
To adjust the zero of the potential, we shall set $\dot \phi_0(0) = -E_{\rm i}/\hbar$ and $\dot \phi_0(t_{\rm f}) = -E_{\rm f}/\hbar$, 
where $E_{\rm i}$ is the energy of the initial state and $E_{\rm f}$ is the one of the final state,
and use an interpolation formula for $\phi_0(t)$ 
with the extra boundary conditions $\phi_0(0)=\phi_0(t_{\rm f})=0$ for simplicity\footnote{Alternatively, we can set the initial and final energy to zero, imposing $\phi_0(t)=0$. This simply amounts to a ``vertical'' shift of the potential with respect to the polynomial  
interpolation in Eq. (\ref{phi_0}). In this paper we use the polynomial form in (\ref{phi_0}).}.
For example,  polynomial interpolations satisfying the boundary conditions for $\eta(t)$ and $\phi_0(t)$ are
\begin{eqnarray}
&& \eta(t) = \frac{t^3}{t_{\rm f}^3} \left[ 1 + 3 \left( 1-\frac{t}{t_{\rm f}}  \right) + 6 \left( 1-\frac{t}{t_{\rm f}}  \right)^2   \right], 
\label{eta}
\\
&& \phi_0(t) = \frac{t}{t_{\rm f}}\left(   1-\frac{t}{t_{\rm f}} \right) \left[  \frac{(E_{\rm i} + E_{\rm f})t-E_{\rm i}t_{\rm f}}{\hbar} \right].
\label{phi_0}
\end{eqnarray}
Note that due to Eqs. (\ref{eq4}), (\ref{b_c_ru}), and (\ref{phi_0}), the phase $\phi(x,t)$ is zero at initial and final times.  
\section{Examples}
\label{ex}

\subsection{Connecting ground states}
\label{gg}
We consider the transformation in a time $t_{\rm f}$ from the ground state wave function of a one dimensional harmonic potential of angular frequency $\omega_{\rm i}$ to the ground state of a harmonic potential of angular frequency $\omega_{\rm f}=\xi \omega_i$. For $\xi<1$ (resp. $\xi>1$), this transformation corresponds to an expansion (resp. compression). Such a transformation can be carried out in a long time using an adiabatic evolution.
Our fast inverse protocol amounts to bypassing the adiabatic evolution.

\begin{figure}[h!]
\centering
\includegraphics[width=8cm]{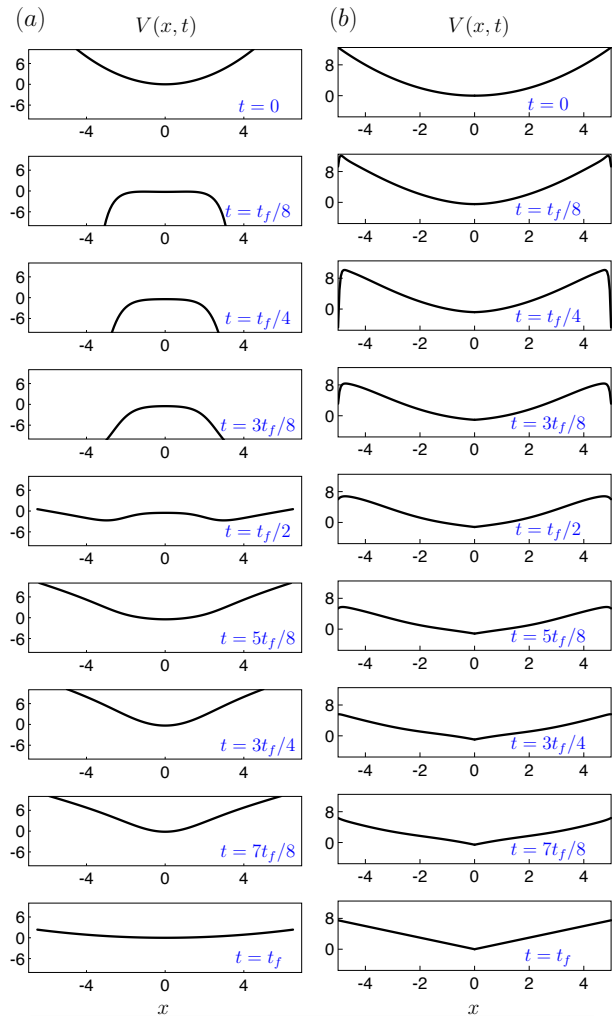}
\caption{(a) Snapshots of the time evolution of the trap shape from a harmonic potential of initial angular frequency $1$ to a harmonic potential of final angular frequency $1/3$ that guarantees the perfect transfer of the ground state of the initial trap to that of the final trap  in a short time $t_{\rm f}=0.24 \times 2\pi$. (b) Snapshots of time evolution of the trap shape from a harmonic potential  $U_H(x)=x^2/2$ to a linear trap  $U_L(x)=3 |x|/2$ in a time interval $t_f=0.24 \times 2\pi$ that ensure the perfect transfer of the ground state of the initial trap to that of the final trap.}
\label{fig1}
\end{figure}

The initial wave function and the final one in dimensionless units ($\hbar=m=\omega_{\rm i}=1$), read 
\beqa
\psi_{\rm i}(x)&=&\pi^{-1/4}e^{-x^2/2}, \nonumber \\
\psi_{\rm f}(x)&=&\xi^{1/4}\psi_{\rm i}(x\sqrt{\xi}).  
\eeqa
The explicit form for ${\cal N}(t)$ and $\rho(x,t)$ can be readily worked out from Eqs. (\ref{rho_p}) and (\ref{eta}). In Fig. \ref{fig1} (a), we provide an example of fast time evolution of the potential $V(x,t)$ from Eq.~(\ref{eqpot}), with $\xi=1/3$ and  $t_{\rm f}=0.24 \times 2\pi$. Note that the transformation is here performed on a timescale significantly smaller than the final period. The curvature of the potential becomes transiently negative at the center of the trap in order to speed up the transformation \cite{PRL10,boltz14}. 
%
%
%
%
%
%

The method developed here also gives the potential to connect two ground states of traps of different kind. This is illustrated in Fig.~\ref{fig1} (b). The ground state of a harmonic potential $U_H(x)=x^2/2$ is transformed into the ground state of a linear potential $U_L(x)=3 |x|/2$ in a time interval $t_{\rm f}=0.24 \times 2\pi$.
\subsection{Splitting of a wave function}
\begin{figure}[h!]
\centering
\includegraphics[width=8cm]{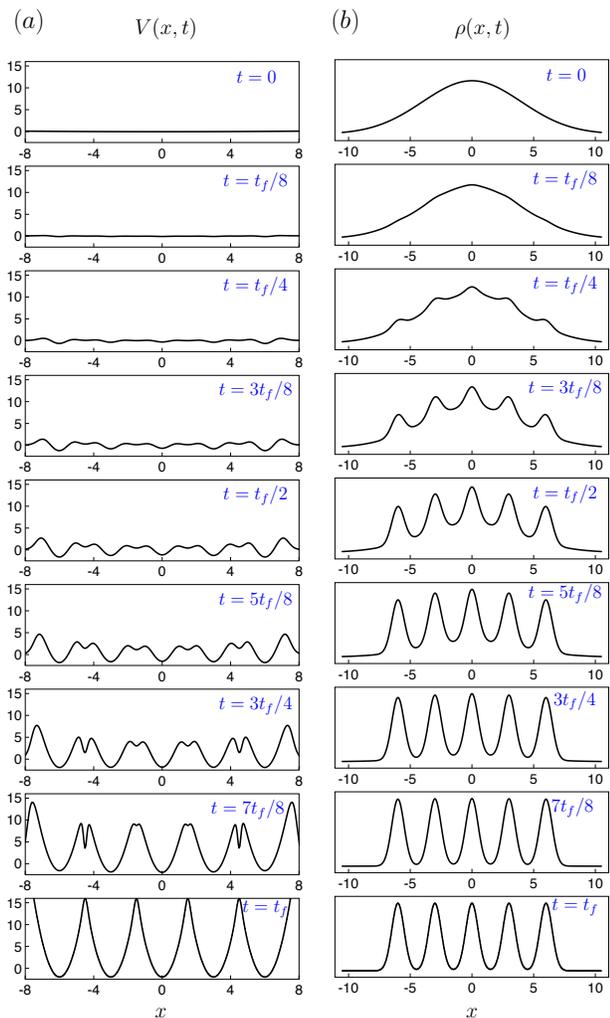}
\caption{(a) Snapshots of the time evolution of the trap shape that splits the wave function of a harmonic potential (of angular frequency $\omega_0=1$) in five equidistant wave functions with the same weight. The transformation is performed in a time $t_{\rm f}=10\pi$. (b) Snapshots of $\rho(x,t)$ 
(arbitrary units) versus position 
in the course of the transformation.}
\label{fig3}
\end{figure}

The invariant-based method and the transitionless tracking approach are problematic for splitting wave functions \cite{torron13}. This is to be contrasted with the simplified fast forward method  \cite{torron13}. The compact formula (\ref{eqpot}) allows for the possibility to split the wave function into an arbitrary number of parts. In Fig.~\ref{fig3} the ground state wave function, $\psi_{\rm i}(x)$, of a harmonic potential of angular frequency $\omega_0$ is split into five parts (given  by ground states of corresponding harmonic oscillators) separated by a distance of $3a_0/4=3(\hbar/m\omega_0)^{1/2}/4$, each having an eighth of the initial width. 
The time-dependent potential deduced from the theoretical framework provides an exact solution for the perfect loading of a periodic structure, an important operation for cold atoms \cite{DSH02,ST02,IR05,ZD09,MNC14,MR14}. The transformation is performed here in a time interval equal to $10\pi$.
for $\omega_0=1$. 
%
%
%
%
%
%
%
%
%
\subsection{Connecting first excited states}
\begin{figure}[h!]
\centering
\includegraphics[width=8cm]{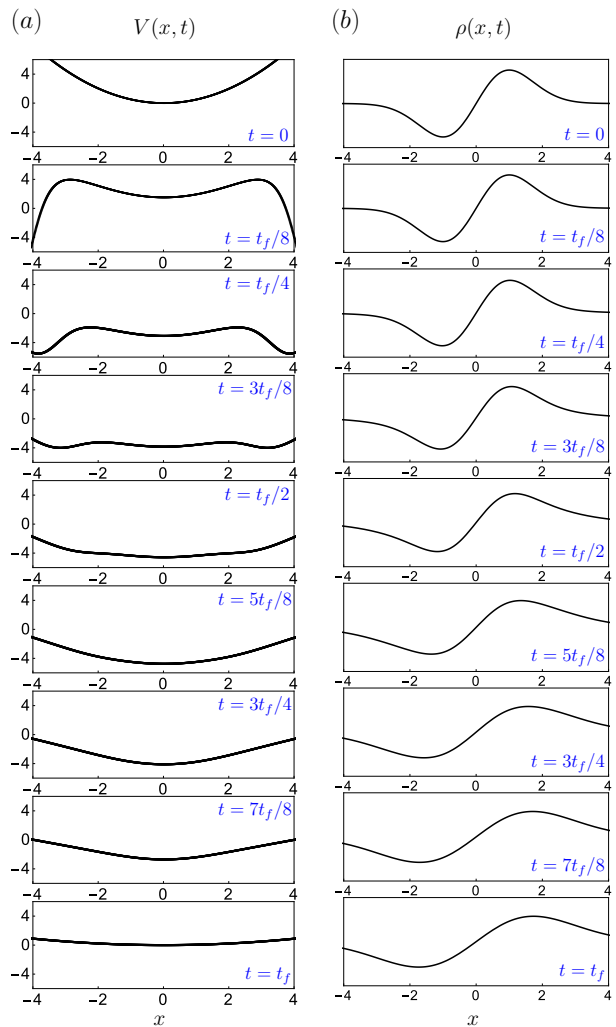}
\caption{(a) Snapshots of the time evolution of the trap shape from a harmonic potential of initial angular frequency $\omega_{\rm i}=1$ to a harmonic potential of final angular frequency $\omega_{\rm f}=1/3$ that ensures the perfect transfer of the first excited state of the initial trap to that of the final trap  in a short time 
$t_{\rm f}=0.48\times 2 \pi$. (b) Snapshots of the time evolution of the corresponding $\rho(x,t)$ (in arbitrary units).}
\label{fig2}
\end{figure}

The approach presented in Sec. \ref{sec2} can be readily generalized to transform the first excited state of a given trap into the first excited state of another trap in a  short time. 
From Eqs. (\ref{eq4}), (\ref{b_c_ru}), and the boundary conditions $\phi_0(0)=\phi_0(t_{\rm f})=0$,  
the condition for the phase $\phi(x,0)=\phi(x,t_{\rm f})=0$ is fulfilled by construction. Thus, 
a way to define an odd state using Eq. (\ref{m-p}) is 
to assume that $\rho(x,t)$ may take negative values. 
The ansatz in Eq. (\ref{rho_p}) generates a positive $\rho(x,t)$ so we set a new one, 
\begin{equation} 
\rho(x,t) = {\cal N}(t) \left\{  \left[1-\eta(t)\right]\psi_{\rm i}(x) + \eta(t)\psi_{\rm f}(x) \right\}.
\label{rho_n}
\end{equation}
For the transformation between two harmonic traps considered in Sec. \ref{gg},  $(\omega_{\rm i}=1, \omega_{\rm f}=1/3)$, 
Figure~\ref{fig2} (a) shows the evolution of $V(x,t)$ using Eq. (\ref{eqpot}) and Eqs. (\ref{eta}), (\ref{phi_0}) and (\ref{rho_n}). In Fig.~\ref{fig2} (b) $\rho(x,t)$ is represented. 
The process time is $t_{\rm f}=0.48\times 2 \pi$. 
%
%
%
%
%
%
%
%
\subsection{Connecting ground and first excited states}
\label{g-f}
The connection between states with different topological properties is of much interest for quantum computing processes because it allows, for example,
to prepare Fock states by deforming the trap without using laser excitation of internal states \cite{PRL13}. Using sequences of $\pi$-pulses is demanding, as a $N$-phonon Fock state needs very precise $N$ pulses, but fluctuations in intensity, frequency, and timing imperfections give a reduced fidelity \cite{Plenio}. 

We want to drive the system from an even state (ground) to an odd one (first excited) of the harmonic oscillator with frequency $\omega_0/(2\pi)$. 
Whenever $\rho(x,t)$ has a definite symmetry (even or odd with respect to $x=0$), the potential in Eq. (\ref{eqpot}) will have even symmetry. 
The reason lies in Eq. (\ref{eq5}), since for an either even or odd $\rho(x,t)$, the ``hydrodynamic velocity'' $u(x,t)$ is always odd. 
Thus the  potential becomes a sum of even functions and stays even throughout the process. 
Therefore, the parity of the initial state will be preserved, and using the interpolation in Eq. (\ref{rho_p}) for $\rho(x,t)$ the system cannot be driven from the even initial state to the desired odd final state. Note in addition the numerical difficulties because of the discontinuity of $\partial_x \rho(x,t)$ at
zeros of $\rho(x,t)$. 

The connection will be achieved by allowing $\rho(x,t)$ to be asymmetric so that the potential in Eq. (\ref{eqpot}) becomes in general 
a sum of even and odd functions. 
The use of Eq. (\ref{rho_n}) as the interpolation formula is a simple way to get an asymmetric $\rho(x,t)$. The potential is singular when $\rho(x,t)=0$ but, as we will see, these are mild singularities in the sense that they may be numerically handled by truncation. 


An alternative way to design the interpolation of $\rho(x,t)$ would be, instead of using Eqs. (\ref{rho_p}) and (\ref{rho_n}), to design $\rho(x,t)$ directly from the wave function as a positive square root of the density, 
\beq
\label{rho_alter}
\rho(x,t)=\sqrt{|\psi(x,t)|^2}.
\eeq
We have unsuccessfully tested this ansatz trying different interpolations for $\psi(x,t)$.
Choosing the interpolating wave function in the form $\psi(x,t)={\cal N} \left[   (1-\eta)\psi_{\rm i} + i \eta\psi_{\rm f}  \right]$ and substituting in Eq. (\ref{rho_alter}), the modulus,
$\rho(x,t) =|{\cal N}|\left((1-\eta)^2|\psi_{\rm i}|^2 + \eta^2|\psi_{\rm f}|^2\right)^{1/2}$, vanishes only at the boundary of the time interval, 
but the resulting $\rho(x,t)$ has a definite even symmetry. 
We have also tried the interpolation $\psi(x,t)=N(t)\{[1-\eta(t)]\psi_{\rm i}(x)+\eta(t)\psi_{\rm f}(x)\}$, which, substituted in Eq. (\ref{rho_alter}), generates an asymmetric function. 
However the desired final state could not be reached numerically due to  singularities that affect the successive derivatives of $\rho(x,t)$, generated because 
of the definition of $\rho(x,t)$ 
via a modulus.      
Therefore, hereinafter we will use the protocol in Eq. (\ref{rho_n}) to connect the ground and first excited states.

\begin{figure}[h!]
\centering
\includegraphics[width=8cm]{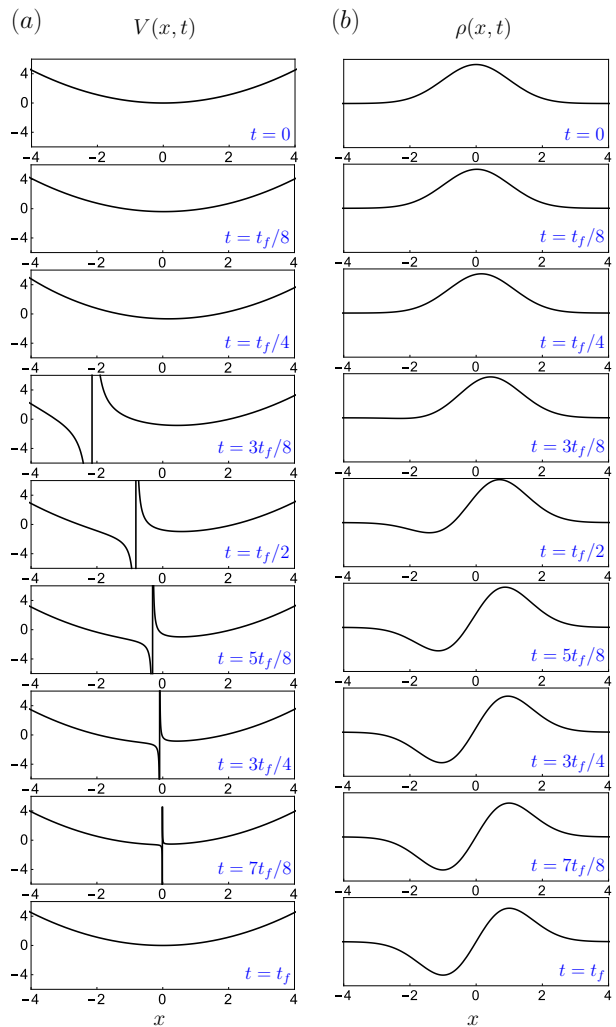}
\caption{(a) Snapshots of the time evolution of the trap shape that connects the ground state to the first excited state of a harmonic potential ($\omega_0=1$). The transformation is calculated in a time $t_{\rm f}=8\pi$. (b) $\rho(x,t)$ (in arbitrary units) is plotted as a function of the position in the course of the transformation.}
\label{fig4}
\end{figure}

Imposing that the initial and final states in dimensionless units ($\hbar=m=\omega_0=1$) are 
\beqa
\psi_{\rm i}(x)&=& \pi^{-1/4} e^{-x^2/2},
\nonumber \\
\psi_{\rm f}(x)&=& \pi^{-1/4} x \sqrt{2} e^{-x^2/2} ,
\eeqa
and substituting in Eq. (\ref{rho_n}), we get the explicit form
\beq
\rho(x,t)=\pi^{-1/4}\frac{1+[x\sqrt{2}-1]\eta(t)}{\sqrt{1+2[\eta(t)-1]\eta(t)}}e^{-x^2/2}.
\label{rho_pi}
\eeq
Introducing Eqs. (\ref{eta}), (\ref{phi_0}) and (\ref{rho_pi}) in Eq. (\ref{eqpot}), we find the potential evolution represented in Fig. \ref{fig4} (a). 
%
%
%
%
%
%
Figure \ref{fig4} (b) shows the evolution of $\rho(x,t)$ given by Eq. (\ref{rho_pi}). 
The system may thus be driven from the ground state into the first excited state without final excitations in a finite, arbitrarily short time. 

\begin{figure}[h!]
\centering
\includegraphics[width=6.cm]{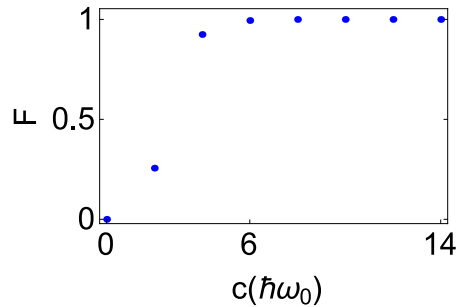}
\caption{Fidelity $F\equiv|\langle \psi^{\rm e}(x,t_f)|\psi_{\rm f}(x) \rangle|^2$ versus the potential truncation constant $c$,  where $\psi^{\rm e}(x,t_f)$ is the final wave function evolved using the truncated potential $V_{\rm trun}$. $t_{\rm f}=8\pi/\omega_0$.}
\label{trun}
\end{figure}

The effect of the divergence of the potential at a position for intermediate times [see Fig. \ref{fig4} (a)] is now studied 
by truncating the potential as
\beq
\label{Vtrun}
V_{\rm trun} (x,t)= \left \{
\begin{array}{lcr}
V(x,t) & \mbox{if } & -c <V(x,t)<c
\\
c & \mbox{if } & V(x,t) \geq c
\\
-c & \mbox{if } & V(x,t) \leq -c,
\end{array}\right.
\eeq
where $c$ is a positive real number.  
To check the stability of the method under this approximation, we compute the fidelity between the final state $\psi_{\rm f}$ and the final state evolved by the truncated potential $\psi^{\rm e}(x,t_{\rm f})$ for different values of $c$.
The evolution of $\psi^{\rm e}(x,t_{\rm f})$ is calculated using the `split-operator method' with the truncated Hamiltonian $H_{\rm trun}=T+V_{\rm trun}$, $T$ being the kinetic energy.
Figure \ref{trun} shows that the method is stable for  large enough $c$.  
For a relative small value of $c=8\hbar\omega_0$ the transition is performed with a $0.9996$ fidelity.

%
%
%
%
%
%
%
\section{Conclusion}
\label{sec4}
We have proposed an improved version of the fast-forward approach described in \cite{torron12,torron13}. In this new formalism, the solution of the time-dependent potential is explicit, so the time necessary to design the potential will be shorter. We applied this technique to accelerate some basic operations that are relevant for quantum information processing and fundamental studies, such as expansions or compressions of a harmonic trap, splitting of a wave function, and generic driving between eigenstates.
In particular, we have studied the connection between the ground and first excited states of a harmonic potential. The connection has been realized allowing  $\rho(x,t)$ to be asymmetric and take negative values. Furthermore, this protocol may also be used to create higher Fock states just by changing the final states.
Open questions left for future work include comparing the present protocol with other methods that break the parity symmetry of the potential 
without using the fast-forward approach \cite{PRL13}, or optimizing the robustness versus noise and perturbations \cite{NJP2012}.
Applications of the method go beyond quantum mechanics, e.g. to determine potentials in a Fokker-Planck equation \cite{Brownian}. 
%
%
%
%
%
\section*{Acknowledgments}
This work was supported by 
the Basque Country Government (Grants No.
IT472-10); 
MINECO (Grant No.
FIS2015-67161-P);  the program UFI 11/55; and by Programme Investissements d'Avenir under the program ANR-11-IDEX-0002-02, reference ANR-10-LABX-0037-NEXT.
M.-G. and M. P. acknowledge fellowships by UPV/EHU. 

\end{document}